\documentclass{iaus}
\usepackage{graphicx}
\usepackage{multirow}

\title[Gaseous Flows in Galaxies] 
{Gaseous Flows in Galaxies}

\author[F. Combes] 
{Francoise Combes$^1$}

\affiliation{$^1$Observatoire de Paris, LERMA (CNRS:UMR8112), 61 Av. de l'Observatoire, 75014 Paris, France
\break email: francoise.combes@obspm.fr}

\pubyear{2007}
\volume{245}  
\pagerange{1--10}
\date{August 2007 and in revised form September 2007}
\jname{Formation and Evolution of Galaxy Bulges}
\editors{M. Bureau, E. Athanassoula, B. Barbuy, eds.}
\begin{document}

\maketitle

\begin{abstract}
The gas component plays a major role in the dynamics of spiral galaxies,
because of its dissipative character, and its ability to exchange angular
momentum with stars in the disk. Due to its small velocity dispersion,
it triggers gravitational instabilities, and the corresponding 
non-axisymmetric patterns produce gravity torques, which mediate these
angular momentum exchanges. When a srong bar pattern develops with
the same pattern speed all over the disk, only gas inside corotation can
flow towards the center.  But strong bars are not long lived in presence
of gas, and multiple-speed spiral patterns can develop between bar phases, 
and help the galaxy to accrete external gas flowing from cosmic filaments.  
The gas is then intermittently driven to the galaxy center, to form 
nuclear starbursts and fuel an active nucleus. The various time-scales of
these gaseous flows are described.
\end{abstract}

\firstsection 

\section{Evidence of gas flows and secular evolution}
 A large wealth of data has recently accumulated, showing the
importance of gas flow in the center, as part of secular evolution
of galaxies. The comparison between the stellar profiles at 3.6 $\mu$m
and the dust profiles at 8$\mu$m with Spitzer, reveals that 
barred galaxies have an excess of gas mass towards the center,
with respect to the extrapolation of the exponential profile (Regan et al 2006).
When the PAH emission is taken as a tracer of star formation rate,
clear differences are revealed in the star formation modes between
pseudobulges, identified by nuclear spirals or bars and flattening of the bulge,
and classical bulge galaxies, with round isophotes.  The latter possess
 star-forming outer disk, with a decline in star formation rates toward the center of the galaxy. 
Star formation is more spectacular in the center of pseudo-bulges (r $<$ 1.5kpc,
Fisher 2006).
All galaxies with a central hole in their PAH (8$\mu$m) distribution are unbarred.
This is expected since, without significant gas inflow,  the central region
of an exponential disk should consume its gas fuel much faster than the outer
parts, according to the non-linear Schmidt law.  

It was already established through CO emission that 
barred galaxies had enhanced central molecular  
gas concentrations, with respect to unbarred galaxies. 
(Sakamoto et al 1999, Sheth et al 2005),
 Now the observation of the same phenomenon in dust
PAH emission at 8 microns consolidates these previous findings,
and in particular indicates that this is not only due to 
an enhanced CO-to-H$_2$ conversion ratio.

Another obvious evidence of gas flows is the existence of
resonant rings in barred galaxies, where the gas is piling up,
and maintained there by the gravity torques of the bar
(e.g. Buta \& Combes 1996). The nature of these rings
and their indication of the bar pattern speed is clear
in galaxies with multiple-rings,
corresponding to the inner or outer resonances.
The PAH dust emission traces beautifully these rings,
previously highlighted by star formation hot
spots, or concentration of CO emission.

\section{Angular momentum transfers}

Originally, the angular momentum (AM)  is created by tidal torques between
structures at early times, before turn-around. The amount
of AM (or J) produced can be quantified by the dimensionless quantity:
$ \lambda = J |E|^{1/2}/GM^{5/2} \sim$ 0.035, where
E is the energy and M the mass.  At this stage, baryons and dark matter 
behave similarly, and they get the same specific angular momentum.
But after turn-around, baryons dissipate and collapse, deeper inside dark haloes,
and they lose AM at the benefit of the dark matter.

In simulation of galaxy mergers, AM is exchanged from baryons to the dark haloes,
through tidal forces and dynamical friction.  The visible component
have then not enough AM to account for the observations.
The dark haloes of merger remnants do not accumulate 
angular momentum: the gravitational heating due to the merger produces
mass loss of the dark matter halo through the virial radius
(D'Onghia \& Navarro 2007).

External gas infall could help to solve the AM problem
of the standard CDM scenario.
Only diffuse cold accretion after the last merger can reform
an extended disk, and this accretion produces secular evolution,
through gas flows.

In purely stellar disks, the secular evolution also
exchanges AM  from baryons to dark haloes; this is a 
way to enhance the formation of bars (Athanassoula 2002, 2003).
If the AM transfer to the dark matter  is saturated, due to an already rotating halo, 
AM is exchanged with the outer disk. The actual mode of transfer
depends on the amount of dark matter present inside the visible disk.
 The AM exchange with the dark matter is made
through resonances, the maximum efficiency being
at corotation. When a bar is present in the disk, 
its pattern speed continuously decreases with time,
due to dynamical friction on the dak halo, which means that 
the corotation resonance moves, and involves always new particles,
that are progressively heated (Debattista \& Sellwood 2000).

\begin{figure}
\begin{center}
\includegraphics[width=13cm]{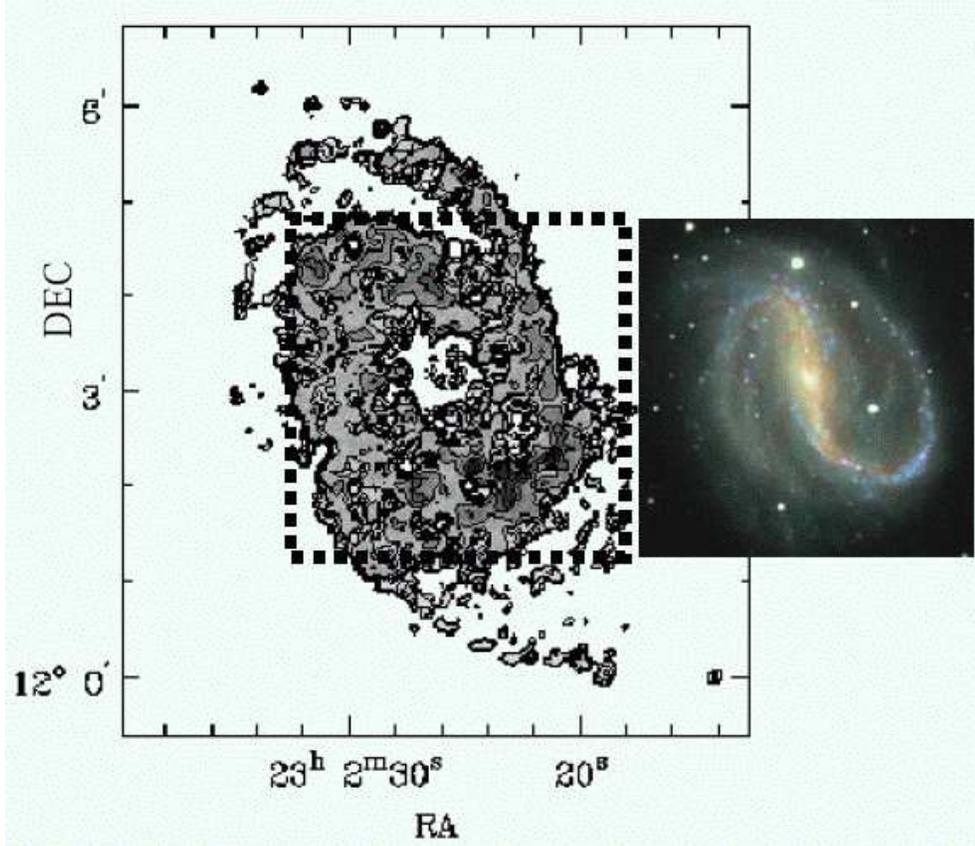}
\end{center}
  \caption{The two spiral patterns in NGC 7479:
at the same scale, are compared an optical image,
which should be overlaid in the dashed square, and
the HI map from Laine et al. (1998a). Note that the optical
spiral arms, getting out of the central bar, are winding up
into a pseudo-ring, which is an outer Lindblad resonance (OLR).
 The inner structure is devoid of atomic gas, while it is rich
in molecular gas, aligned along the bar (Laine  et al 1999).
 The HI emission begins in a ring, at the OLR of the optical
pattern, and then continues in an external spiral structure.}
\label{n7479-tot}
\end{figure}

When the galaxy is rich in gas, the angular momentum exchange
is preferentially done between stars and gas, given the strong
torques exerted from the non axisymmetric stellar patterns to
the dissipative and density-contrasted gas component.
Viscous torques are also present at a very low level,
but are easily dominated by the gravity torques
(Bournaud \& Combes 2002). Gravity torques change sign
at each resonance, they drive the gas inwards from corotation
to the center. In this radial gas inflow, the 
AM lost by the gas is given to the bar. It is comparable in amplitude
to the AM of the bar wave, and contributes to weaken or destroy it. 
This rate of exchange has been  quantified by observations:
torques can be computed with the help of the near
infrared images, as tracer of the mass distribution, 
and with the gas distribution (H$\alpha$, CO, HI). The torque
is proportional to the phase shift between the stellar bar and
gas response, and the leading dust lanes characteristics of bars
are already an obvious smoking gun (e.g. Garcia-Burillo et al 2005).

The gravity torques are positive outside corotation,
and the gas is flowing outwards until the outer Lindblad resonance (OLR).
Does it mean that gas in the outer disks encounter
difficulties to flow in?   This might be the case when the bar
is strong, and forces its pattern speed to the outer spiral.
But in the case of a weak pattern, simulations show that
the most frequent dynamical
state of a galaxy is with multiple pattern speeds.
 The outer disk decouples from the center, and develops a 
spiral at lower pattern speed. This could be the case of many 
extended gaseous disks, which correspond sometimes to 
extended UV disks. The activity in the outer parts
could be triggered by companions.
An example is given by NGC 7479
in Fig. \ref{n7479-tot}, where the atomic gas has a morphology
apparently decoupled from the thin central bar.

\begin{figure}
\begin{center}
\includegraphics[angle=-90,width=11cm]{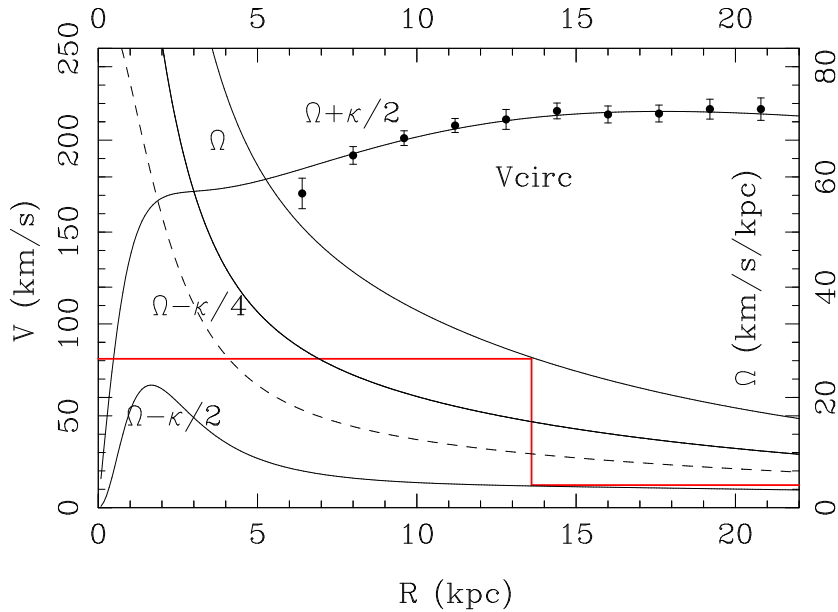}
\end{center}
  \caption{The rotational velocity in NGC 7479 (left-scale),
and corresponding frequencies (right scale).
 The data points are from the HI kinematics (Laine et al 1998a),
and the model circular velocity has been fitted to them.
In the central part, where the CO gas has strong non-circular
motions, the model comes from the gravitational potential
derived from the near-infrared image (Block et al 2002). 
The central pattern speed of 27km/s/kpc
has been estimated by Laine et al. (1998b). The optical
spiral pattern is outside corotation, and winds up at OLR.
We propose here that this ring, which corresponds also to
an HI ring, is the ILR of the external spiral pattern,
with speed 4km/s/kpc.}
\label{n7479-vcur}
\end{figure}

The examination of the gravitational potential and rotation curve
of NGC 7479 reveals that the pseudo-ring in the optical (and HI)
might be the OLR of the central bar, and the ILR of the outer
spiral pattern, as shown in Fig. \ref{n7479-vcur}.

\section{Bar destruction, role of gas and dark matter}

For more than a decade now, numerical simulations
of barred galaxies with gas have revealed the fragility of bars,
which can be easily detroyed by secular evolution
(e.g. Friedli \& Benz 1993). But this destruction depends
on several factors, and some controversy has been discussed
in the literature.   Since the gas inflow towards the center is accompanied
by a central mass concentration, the latter has been first thought
to tbe the main destruction mechanism.  But a central mass concentration
alone must be very high to succeed to weaken the bar
(Shen \& Sellwood 2004, Athanassoula et al 2005).

The main destruction mechanism could be 
the  gas inflow itself, which is driven in by the bar torques.
The angular momentum is taken up  by the bar wave
(e.g. Bournaud et al 2005a). When the galactic disk is not
dominated by dark matter, the AM exchange available for the  stellar
component is only the outer disk or the gas. The outer disk takes the
AM at the bar formation. When the gas inflows, it gives back
AM to the bar, and weakens it, the
mechanism can be called self-destruction of the bar.

The actual effect of the gas depends on the cooling, and is
very different when the gas is considered isothermal or adiabatic 
(Debattista et al. 2006). In presence of cooling, the gas is very dissipative,
which enhances its phase-shifted reponse and the amplitude of the 
torques: the bar destruction is then quite rapid.

Let us note that the AM exchange redistributes
mass radially, and can explain exponential profiles.
The succession of several bars could have signatures in
density profiles breaks, that  are now currently observed
in may be 75\% of galaxies (Pohlen 2002).

\begin{figure}
\begin{center}
\includegraphics[angle=-90,width=12cm]{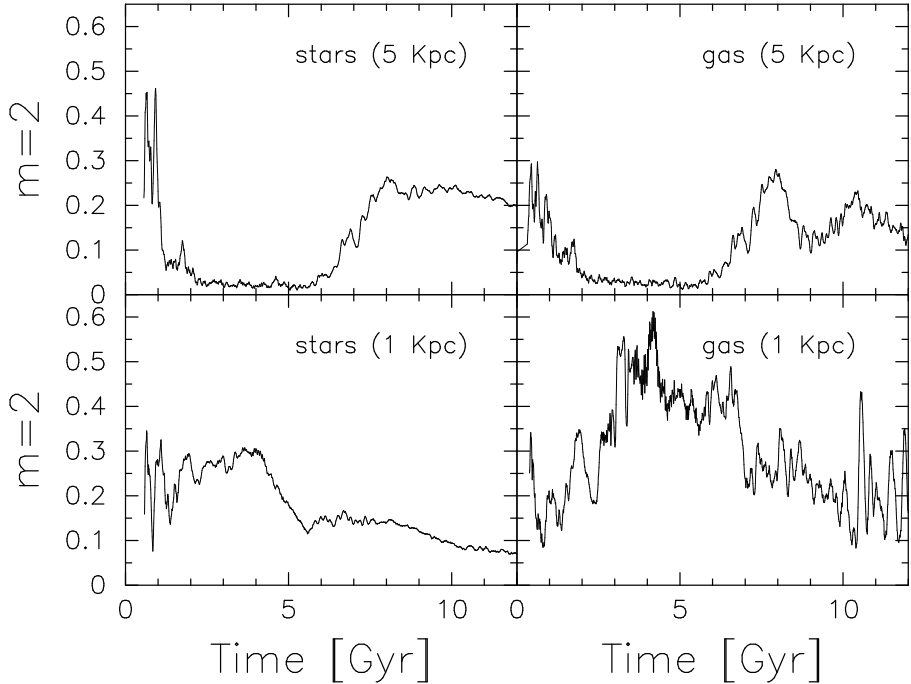}
\end{center} 
 \caption{The evolution of a galaxy disk, inside a triaxial dark matter
halo collapsed at z=2. The $m=2$ Fourier component is shown versus time for
stars (left frames) and gas (right frames) within the central 5~kpc (upper frames)
and 1~kpc (lower), from Heller et al. (2007b).}
\label{heller07}
\end{figure}

The bar destruction in presence of large gas fraction
has also been seen in simulations of galaxy
formation in a cosmological context, even in presence
of massive triaxial dark matter haloes
(Heller et al 2007b).
Bars form and drive gas inflows towards the center,
which trigger formation of nuclear bars;  this in turn
weakens and destroys the primary bars.
The amplitude of the $m=2$ component either in 
the stellar or gaseous component has a complex
evolution (see Fig \ref{heller07}).

Bars can reform in galactic disks, 
through external gas accretion, and secular evolution.
 But the inflow of external gas has to wait the 
weakening of bars and their corresponding 
positive torques outside corotation, which prevent
the gas enter.
In that sense, gas accretion is intermittent;
on a time-scale of several hundreds million years,
the gas is first confined outside OLR until the bar weakens,
 then it can replenish the disk, to make it unstable again to bar formation.
 The gas inflow towards the very center, to feed AGN, is also
intermittent; the gas in a first phase is stalled at the ILR,
and has to wait for the decoupling of a nuclear bar, to be
driven further inwards. This final steps are quite rapid,
of the order of the dynamical time-scales at these radii,
which are of the order of 10 Myr.

Different conclusions have been drawn recently by
Berentzen et al (2007), with simulations of
galaxies embedded in massive dark haloes.
 They compute the exchange of AM between stars and gas,
and find it negligible, the AM exchange being essentially
with the dark matter halo particles.
 Their gas fraction is always smaller than 8\%, may be insufficient
for the gas to have a significant role in the AM transfer in presence
of massive haloes.

However, they find that bars weaken and destroy more quickly in gas-rich galaxy
disks, than in gas-poor ones, the separating gas fraction being 3\%.
They interprete this phenomenon in terms of central mass concentration.
The gas driven inwards by the bar creates a central mass concentration,
which weakens the bar, in destroying the resonances and heating the stars.
  In gas poor galaxies, the bar lasts longer, and the main weakening 
mechanism for the bar is the peanut formation, through vertical resonances.
  Gas-rich galaxies have little or no peanut instability, because the central
mass concentration damps it. One common conclusion, nevertheless, is that bars
disappear more quickly in presence of gas.

\begin{figure}
\begin{tabular}{c}
\includegraphics[width=6cm]{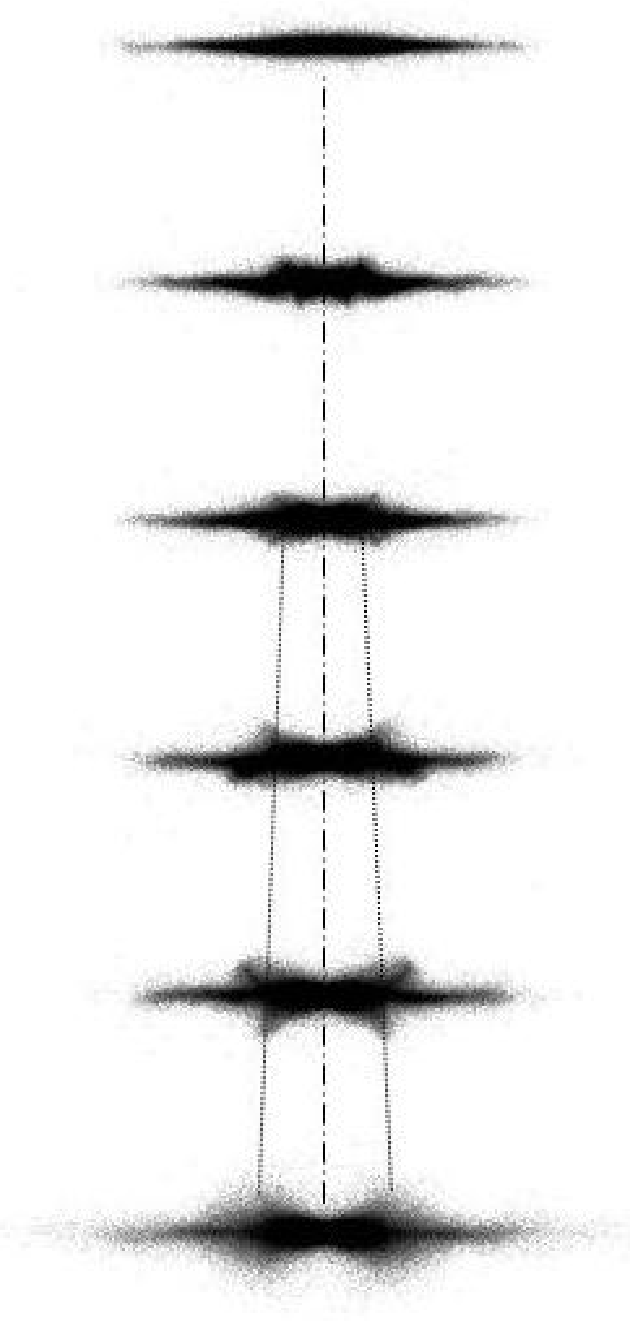}\\
\end{tabular}
\begin{tabular}{c}
  \includegraphics[width=7cm]{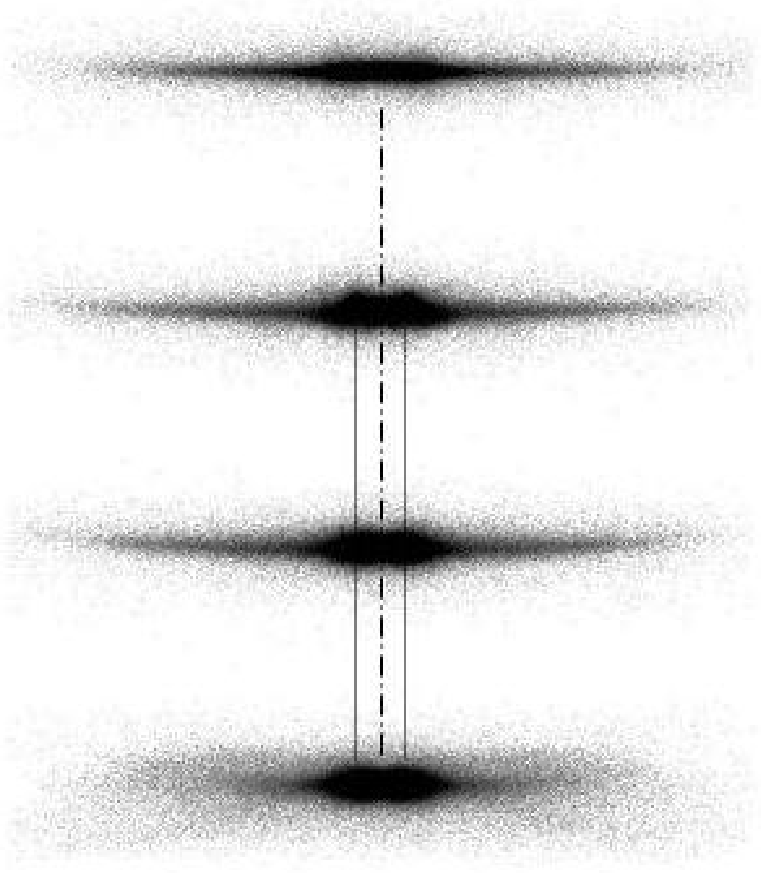}\\
  \includegraphics[width=7cm]{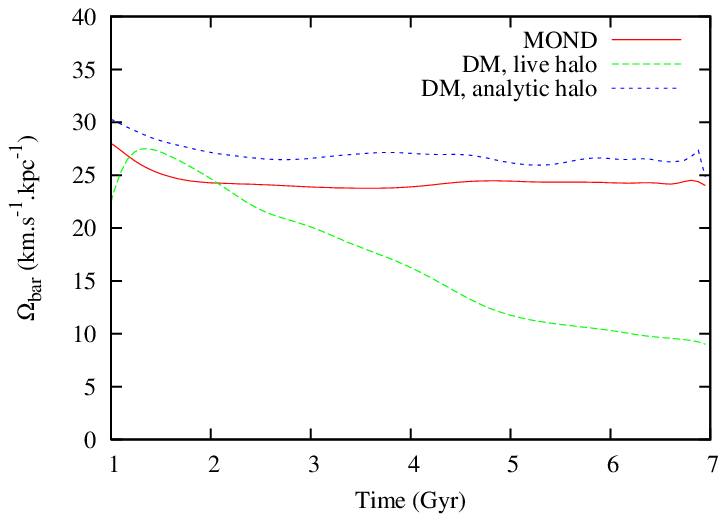}\\
\end{tabular}
  \caption{ Left: peanut formation in a galaxy embedded in a 
dark matter halo, note the increase of the peanut radius with time.
Top right: buckling instability in a MOND galaxy, the peanut
radius is now fixed. 
Bottom right: Evolution with time of the bar pattern speed, in the 
case of a live dark matter halo, an analytical halo, or MOND.
In both latter cases, the bar does not experience dynamical friction,
and keeps a constant speed, from Tiret \& Combes (2007).}
\label{peanut}
\end{figure}

The precise role of the dark matter halo can be quantified,
by comparing for instance the dynamics of galaxies, treated
with DM or without, but in modified gravity, so that they have the
same rotation curves (e.g. Tiret \& Combes 2007).
With a dark matter halo, the bar appears later, the disk
being stabilised by it. In the MOND case, the disk is entirely
self-gravitating and more unstable.  For pure stellar disks
without gas, the destruction of the bar occurs when the
peanut instability thickens the disk vertically.
The buckling instability occurs in both cases, but
later in MOND. The main differences between the two
models is the existence of the dynamical friction against 
the dark matter halo, and the exchange of angular momentum.
  Because of the friction, the pattern speed declines in the
DM case, while it keeps constant in MOND (cf Figure \ref{peanut}).
 The resonance shifts in radius, and the peanut instability
also. Through AM exchange with the halo, the bar
can reform after the peanut weakening.

\begin{figure}
\begin{center}
\includegraphics[angle=-90,width=11cm]{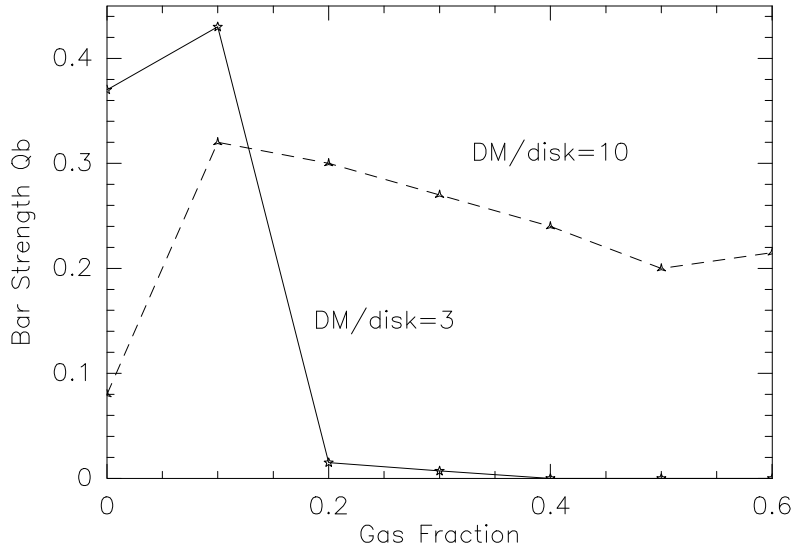}
\end{center}
  \caption{ Bar strength Q$_b$ of cosmologically evolved galaxies,
versus their gas fraction,  for two mass ratios between dark matter halo and
baryonic disk: DM/disk = 3 (full line) and 10 (dash), from Curir et al (2007).}
\label{curir}
\end{figure}

As for the role of gas, it has been shown also in  
cosmological simulations, that the bar strength declines faster
in the presence of gas, but this is more visible when the dark
matter halo is not dominating.
The fraction of gas able to destroy the bar depends on the DM/disk 
mass ratio, it is f$_{gas}$ =0.2, when DM/disk = 3  (Curir et al. 2007,
see Figure \ref{curir}).
When the DM halo mass is negligible within the disk, this fraction falls
down to 6\%.

Heller et al (2007a) have investigated the formation of
nested bars in the early evolution.
The secondary bar lasts only 1-2 Gyr, and is
destroyed after a large gas inflow to the center.
In the same time, the primary bar weakens.

\begin{figure}
\begin{center}
\includegraphics[angle=-90,width=13cm]{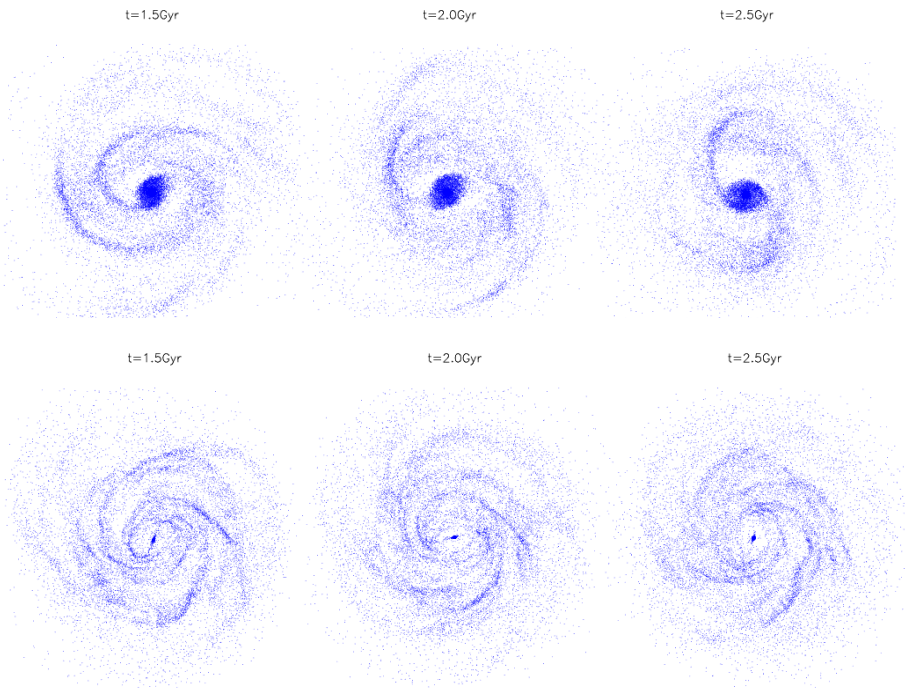}
\end{center}
  \caption{ The evolution of the gas component depends on
the shape of the dark matter halo, for the same total masss. 
{\it Top:} simulation in an NFW dark matter live halo, compared
to {\it Bottom:} in a Plummer live halo. The simulations
have been done with a Tree-SPH code, described
in di Matteo et al (2007a).}
\label{nfw-plum}
\end{figure}

The dark matter halo plays a large role in the
development of bar instability, and therefore gas flows
in galaxy disks, but not only through its total mass,
also through its radial profile.
Figure \ref{nfw-plum} shows the gas distribution
in disks embedded in DM haloes of the same mass, but one
has an NFW profile, while the other is a Plummer.
  In the NFW case, the DM in the center is cuspy,
which creates an axisymmetric mass concentration,
and dilutes the gravity torques of the bar: the gas flows are
hampered, while in the Plummer case, the gas flows
towards the center, and develops the classical ringed/bar 
morphology.

\section{Cold gas accretion}

The majority of galaxies are barred (75\%)
and most of them (80\%) are strongly barred, as found
in the near infrared (Eskridge et al 2002, Block et al 2002).
There is in particular a paucity of weak bars (Marinova \& Jogee 2007).
The bar frequency appears constant with redshift
(Jogee et al 2005).
Since bars are not long-lived in gaseous disks, the observed
bar frequency in late-type galaxies require that secular evolution
reform bars. This can be done through external gas accretion,
which replenishes the disk and re-creates the conditions of bar 
instability, but not with galaxy interactions, which tend to
heat and destroy thin disks.
 
Cold gas accretion can also explain the large frequency of
lopsided galaxies, and in particular in 
late-types (77\% are asymmetric, Matthews et al 1998).
Even in the near-infrared, about 20\% of galaxies
have their Fourier component A$_1 >$ 0.2. 
Lopsidedness can be produced in interactions,
but is not long-lived, and an external mechanism
is required (Bournaud et al. 2005b).
Asymmetric gas accretion (with $\sim$ 4 M$_\odot$/yr)
can explain the observed frequency of $m=1$ perturbations
and their long life-time. 

Cold gas accretion is also required to reform disks after mergers
and also to account for star formation histories that are
not falling off exponentially with time, in intermediate-type spirals.
Cosmological accretion can typically double the mass of
a galaxy in 10Gyr.
Cold gas accretion helps to solve the angular momentum
problem (D'Onghia et al 2006).

\section{AGN fueling and feedback}

Gas radial inflows are part of a dynamical
cycle involving bars, they can feed AGN or starburst,
that are themselves providing negative feedback to stop
their activity.  Gas flows are then intermittent.
The time-scale corresponding to each phase can
be estimated from the statistical frequency of
galaxy observations in the precise phase.

From a detailed study of the molecular gas in the circumnuclear
regions of active galaxies (NUGA project), it appears that 
the AGN fueling phase must be very short, corresponding 
to the dynamical time-scale in the central 100-200pc.
 In the very center, the gas is sometimes
driven inwards through nuclear spirals,
conspicuous from their dust absorption,
but also underlined by CO emission. These nuclear
spirals could be driven by non-axisymmetric patterns or by
orbiting black holes (Etherington \& Maciejewski 2007).

Once the AGN activity is started, the energy released,
and in particular the radio jets, provide a negative feedback
to stop further gas accretion. 
The AGN feedback is however able to influence the galaxy as a whole
only for large masses, and in particular in massive galaxies
in the center of cooling flows. It can explain 
the exponential cut-off at the bright end 
of the galaxy luminosity function (Croton et al 2006).

The exploitation of this feedback mechanism in cosmological
simulations, allows to understand better the correlation
and concomittant growth of spheroids and black holes,
the AGN activity influencing and moderating the star formation
activity of their hosts. Following their growth
as a function of redshifts and along the merger hierarchy, it can be found that
most of supermassive black holes accrete their mass in the quasar regime, 
driven by mergers. The quietest way (much lower than the Eddington rate)
 is marginal (Sijacki et al 2007).

\section{Gas flows in interactions and mergers}

Important gas flows are triggered by galaxy interactions and
mergers. They are also driven by gravity torques,
produced by the non-axisymmetry of the stellar component,
through bars essentially.

A recent series of merger simulations, varying the geometry
of the galaxy interactions, the morphological types and mass
ratios, has shown that gas inflows, and then the corresponding 
starbursts, were more efficient for retrograde encounters
(di Matteo et al 2007a). This might appear paradoxical,
but can be explained when the sense of gas flows are
computed in detail. In direct interactions, the tides
are more resonant, and gas is efficiently driven outwards,
in tidal tails. In retrograde encounters, on the contrary,
the gas remains in the disks, and is inflowing due to bars.

The angular momentum exchanges are from the initial relative
orbit to the dark matter halo essentially, but also to the
visible disk, gas or stars. Some cases of retrograde encounters
between an elliptical and a spiral galaxies,
reveal the formation of counter-rotating cores,
by a new mechanism (di Matteo et al 2007b).
Since tidal forces are stronger in the outer parts,
the disk of the spiral is highly perturbed, and 
absorbs a large part of the negative relative
orbital momentum.
Its center is unaffected, and keeps its initial spin orientation.
The stellar component from the elliptical, that was not
rotating initially, also absorbs part of the orbital AM,
as the dark matter. The counter-rotating effect is
opposite to the mechanism described in 
Balcells \& Gonzalez (1998).

\section{Conclusion}

Secular evolution plays a fundamental role, in the fueling of 
starbursts and AGN, and in bulge formation.
 The bar instability is the main motor of this evolution,
and is part of a dynamical cycle, leading to its
own destruction. Starting a new cycle and reforming a bar
requires diffuse cold gas accretion, from cosmic filaments,
replenishing the galactic disk in gas.

Along the cycle, the stars in the disk exchange angular momentum with dark matter 
haloes or with the outer disk, or with gas,
according to dark-to-visible mass ratio, saturation or resonances.
The gas inflow itself leads to the decline in bar strength.
The gas fraction able to destroy bars depend on the DM/disk mass ratio.

In addition to the total amount of dark matter halo, 
gas flows and bar strength depend on dark halo radial profiles,
orbital chaos and resonances, gas fraction,
vertical feedback (peanut), etc.. 

Bars trigger gas flows in galaxy interactions and mergers.
The gas inflows and the subsequent starbursts are 
more efficient with retrograde orbits,
since the tidal tails dragging the gas
out of the galaxies are then less developped.


\begin{discussion}

\discuss{Massey}{I'm wondering .................
dispersion in M$_v$?}

\end{discussion}

\end{document}